\def\w0{\omega_0}
\documentclass{dcds}
\usepackage{epsfig}
\usepackage[dvips]{color}

\def\currentvolume{}
 \def\currentissue{}
  \def\currentyear{}
   \def\currentmonth{}
    \def\ppages{--}

\title[Noise in ecosystems:
 a short review] {Noise in ecosystems: A short review}
\author[B. Spagnolo, D. Valenti, A. Fiasconaro]{}

\thanks{Research supported by INTAS Grant
01-450, by MIUR and INFM}
 \email{spagnolo@unipa.it}

\begin{document}
\maketitle

\centerline{\scshape  B. Spagnolo, D. Valenti, A. Fiasconaro}
\medskip

{\footnotesize \centerline{Dipartimento di Fisica e Tecnologie
Relative} \centerline{Istituto Nazionale di Fisica della Materia,
Unit\`a di Palermo}  \centerline{Universit\`a di
Palermo}\centerline{Viale delle Scienze, I-90128 Palermo, Italy}}
\medskip



\centerline{(Communicated by  Stefano Boccaletti)}

\bigskip

\begin{quote}{\normalfont\fontsize{8}{10}\selectfont
{\bfseries Abstract.} Noise, through its interaction with the
nonlinearity of the living systems, can give rise to
counter-intuitive phenomena such as stochastic resonance,
noise-delayed extinction, temporal oscillations, and spatial
patterns. In this paper we briefly review the noise-induced
effects in three different ecosystems: (i) two competing species;
(ii) three interacting species, one predator and two preys, and
(iii) N-interacting species. The transient dynamics of these
ecosystems are analyzed through generalized Lotka-Volterra
equations in the presence of multiplicative noise, which models
the interaction between the species and the environment. The
interaction parameter between the species is random in cases (i)
and (iii), and a periodical function, which accounts for the
environmental temperature, in case (ii). We find noise-induced
phenomena such as quasi-deterministic oscillations, stochastic
resonance, noise-delayed extinction, and noise-induced pattern
formation with nonmonotonic behaviors of patterns areas and of the
density correlation as a function of the multiplicative noise
intensity. The asymptotic behavior of the time average of the
\emph{$i^{th}$} population when the ecosystem is composed of a
great number of interacting species is obtained and the effect of
the noise on
the asymptotic probability distributions of the populations is discussed.\\
\par}
\end{quote}

\section{Introduction}\label{S:1}
In recent years several theoretical investigations have been done
on noise-induced effects in population
dynamics~\cite{Ciu}-\cite{CirPasSpa}. In particular, the problem
of the stability of complex ecological systems in the presence of
noise has been widely discussed~\cite{Sci99}. New counterintuitive
phenomena, such as stochastic resonance~\cite{Gam,Man}, noise
enhanced stability~\cite{AguSpa} and noise delayed
extinction~\cite{Spa1,SpaFiaVal1,ValFiaSpa1}, can appear because
of the presence of noise in living systems, whose dynamics is
nonlinear. The interaction between noise and nonlinear determinism
in ecological dynamics adds an extra level of complexity compared
with the largely stochastic dynamics of, say, economic systems or
the largely deterministic dynamics of many physical and chemical
processes \cite{eco01}. Ecological systems are open systems in
which the interaction
 between
the component parts is nonlinear and the interaction with the
environment is noisy. This intrinsic nonlinearity can give rise to
the complex behavior of the system, which becomes very sensitive
to initial conditions, various deterministic external
perturbations, and to fluctuations always present in nature. The
comprehension of noise's role in the dynamics of nonlinear
 systems plays a
key aspect in the efforts devoted to understand and then to model
so-called complex ecosystems. One approach to understanding the
complexity is to start with a conceptually simple view of the
system in order to catch the phenomena of interest and then to add
details that introduce new levels of complexity
\cite{Sci99,kad99}. In general the effects of small perturbations
and noise, which are ubiquitous in real systems, can be quite
difficult to predict and often yield counterintuitive behavior.
Even low-dimensional systems exhibit a huge variety of
noise-driven phenomena, ranging from a less ordered to a more
ordered system dynamics.

In the past, the study of deterministic mathematical models of
ecosystems has clearly revealed a large variety of phenomena,
ranging from deterministic chaos to the presence of a spatial
organization. These models, however, do not account for the
effects of noise despite the facts that it is always present in
actual population dynamics and that it arises from different
sources, such as the intrinsic stochasticity associated with the
random variability of the environment. Frequently, its effects
have been assumed to be only a source of disorder. Recently
researchers have shown a growing interest in a deeper
understanding of the effects of fluctuations in biological systems
ranging from neuroscience to biological evolution and to
population dynamics \cite{Ciu}-\cite{CirPasSpa},
\cite{eco01}-\cite{tur00}.

In addition, analyses of experimental data of population dynamics
frequently need to consider spatial heterogeneity. Characterizing
the resultant spatio-temporal patterns is, perhaps, the major
challenge for ecological time series analysis and for dynamics
modeling. To describe complex ecosystems, it is therefore
fundamental to understand the interplay between noise, periodic
and random modulations of some environment parameters, and the
intrinsic nonlinearity of simple models of ecosystems and to
understand spatio-temporal dynamics
\cite{Spa1,Spa2},\cite{zho98}-\cite{bal99}.

The principal aim of this work is to review some recent results
obtained for systems described in term of a generalized
Lotka-Volterra model including a term of multiplicative
noise~\cite{Ciu}. A constructive role of the noise is observed. It
contributes to producing: (a) quasi-periodic oscillations and
stochastic resonance in the presence of a driving force; (b)
noise-delayed extinction, (i.e. a nonmonotonic behavior of the
average extinction time of one of the two species as a function of
the noise intensity); and (c) nonmonotonic behavior of the pattern
formation, the density correlation and pattern areas as a function
of noise intensity. We analyze three different ecosystems
described by the formalism of the Lotka-Volterra equations. The
first ecosystem is comprised of two competing species in the
presence of two noise sources: a multiplicative noise which
affects directly the dynamics of the species, and an additive
noise responsible for the random behavior of the interaction
parameter between the species. We obtain quasi-periodic
oscillations of two species densities, stochastic resonance (SR)
and noise-delayed extinction. We also investigate the system using
multiplicative colored noise with different values of the
correlation time $\tau_c$. The effect of the correlated noise is
to shift the peak of the signal-to-noise ratio (SNR), which is the
signature of the SR phenomenon. For this ecosystem we also
analyzed the spatial effects by considering a discrete time
evolution model of the Lotka-Volterra equations with diffusive
terms, namely a coupled map lattice (CML), and we analyzed the
spatio-temporal patterns of the two species induced by the noise.
In the second ecosystem, comprised of three interacting species,
namely one predator and two preys, we analyzed the spatio-temporal
behavior of the species densities. We find: (a) noise-induced
pattern formation in the coexistence regime, which depends on the
initial conditions, (b) oscillating behavior of the site
correlation coefficient with an alternation between coexistence
and exclusion regime, (c) nonmonotonic behavior of the pattern
area as a function of noise intensity. Finally we consider a
system comprised of many interacting species. The analytical
resolution of the Lotka-Volterra equations is more difficult in
the presence of a large number of species. Nevertheless some
analytical approximations for the mean field interaction between
the species as well as numerical simulations give some insight
into the behavior of complex ecosystems~\cite{Ciu,CirPasSpa,Cir1}.
For a large number of interacting species, it is reasonable, as a
phenomenological approach, to choose the growth parameter and the
interaction parameter at random from given probability
distributions. Within this type of representation, the dynamics of
coevolving species can be characterized by statistical properties
over different realizations of parameter sets. Though the
generalized Lotka-Volterra model has been explored in
detail~\cite{SviLog}, it seems that a full characterization,
either deterministic or statistical, of the conditions under which
a population becomes extinguished or survives in the competition
process, has not been achieved~\cite{Abram,AbramZa}. In this last
ecosystem, two types of interaction between the species have been
considered: (a) mean field interaction, and (b) random
interaction. We focused on the statistical properties of the
$i^{th}$ population, obtaining the asymptotic behavior of the time
integral and the distributions both of the population and the
local field, which is the interaction of all species on the
$i^{th}$ population. By introducing an approximation for the time
integral of the average species concentration $M(t)$ we obtained
analytical results for the transient behavior and the asymptotic
statistical properties of the time average  of the \emph{$i^{th}$}
population.

\section{Two competing species}\label{S:2}
Time evolution of two competing species is obtained within the
formalism of the Lotka-Volterra equations~\cite{Lot} in the
presence of a multiplicative noise

\begin{eqnarray}
\frac{dx}{dt}=\mu_1\thinspace x\thinspace(\alpha_1-x-\beta_1(t) y)+x\thinspace\xi_x(t)\\
\frac{dy}{dt}=\mu_2\thinspace y\thinspace(\alpha_2-y-\beta_2(t)
x)+y\thinspace\xi_y(t),
 \label{LotVol}
\end{eqnarray}
where $\xi_x(t)$ and $\xi_y(t)$ are statistically independent
Gaussian white noises with zero mean and correlation function
$\langle \xi_i(t)\xi_j(t')\rangle = \sigma
\delta(t-t')\delta_{ij}$ ($i,j=x,y$).
\begin{figure}[htbp]
\begin{center}
\includegraphics[width=8cm]{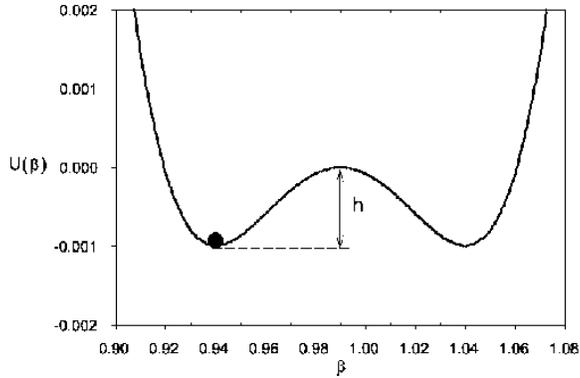}
\caption{The bistable potential $U(\beta)$ of the interaction
parameter $\beta(t)$. The potential $U(\beta)$ is centered on
$\beta=0.99$. The parameters of the potential are $h = 6.25 \cdot
10^{-3}$, $\eta=0.05$, $\rho = -0.01$.} \label{potential}
\end{center}
\end{figure}
It is known that the real biological systems are affected by
random interactions, because of the presence of environmental
fluctuations. The noise and some other deterministic periodical
driving force present in the ecosystems, such as the temperature,
contribute to determine also the dynamics of $\beta$, the
interaction parameter between the species. For $\beta < 1 $ a
coexistence regime takes place (that is, both species survives),
while for $\beta > 1 $ an exclusion regime is established (that
is, one of the two species vanishes after a certain time).
Coexistence and exclusion of one of the two species correspond to
stable states of the Lotka-Volterra's deterministic
model~\cite{Baz}. The change in the competition rate between
exclusion and coexistence occurs randomly because of the coupling
between the limiting resources and the noisy environment. A random
variation of limiting resources produces a random competition
between the species. The noise therefore, together with the
periodic force, determines the crossing from a dynamical regime
($\beta < 1$, coexistence) to the other one ($\beta > 1$,
exclusion)~\cite{Spa2,ValFiaSpa1}. To describe this continuous and
noisy behavior of the interaction parameter $\beta(t)$ we consider
a stochastic differential equation with a bistable potential and a
periodical driving force

\begin{equation}
\frac{d\beta(t)}{dt} = -\frac{dU(\beta)}{d\beta}+\gamma
cos(\omega_0 t) + \xi_{\beta}(t) \label{beta_eq},
\end{equation}
where $U(\beta)$ is a bistable potential (see Fig. 1)

\begin{equation}
U(\beta) = h(\beta-(1+\rho))^4/\eta^4-2h(\beta-(1+\rho))^2/\eta^2,
\label{U(beta)}
\end{equation}
and $h$ is the height of the potential barrier. The periodic term
takes into account for the environment temperature variation. Here
$\gamma=10^{-1}$ and $\omega_0/(2\pi)=10^{-3}$. In Equation
(\ref{beta_eq}) $\xi_{\beta}(t)$ is a Gaussian white noise with
the usual statistical properties: $\langle
\xi_{\beta}(t)\rangle=0$ and $\langle
\xi_{\beta}(t)\xi_{\beta}(t')\rangle =
\sigma_{\beta}\delta(t-t')$. Due to the shape of $U(\beta)$ it is
reasonable to expect a coexistence regime for $\beta(0) < 1 $,
when deterministic case ($\xi_{\beta}(t)=0$) is considered.

\subsection{Stochastic resonance}\label{ss:2.1}

First, we investigate the effect of the noise on the time behavior
of the species.
\begin{figure}[htbp]
\begin{center}
\includegraphics[width=11cm]{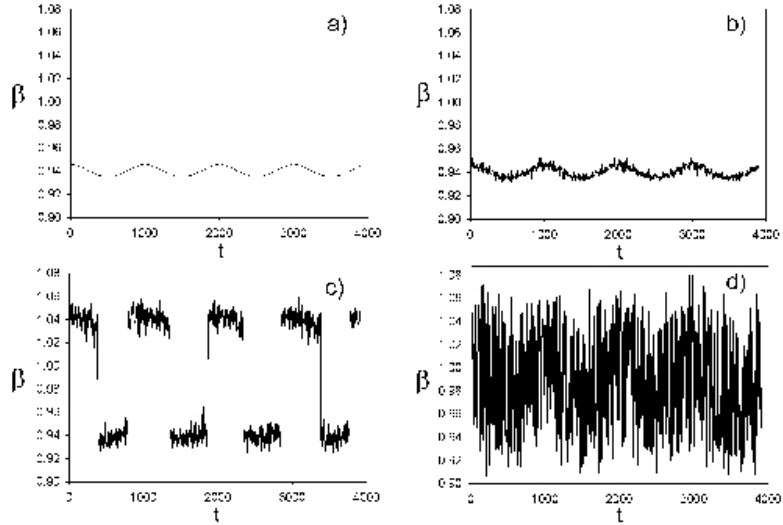}
\caption{Time evolution of the interaction parameter for different
values of the additive noise $\sigma_\beta$. (a) $\sigma_\beta=0$;
(b) $\sigma_\beta=1.78\cdot 10^{-4}$; (c) $\sigma_\beta=1.78\cdot
10^{-3}$; (d) $\sigma_\beta=1.78\cdot 10^{-2}$. The values of the
parameters are: $\gamma=10^{-1}$, $\omega_0/(2\pi)=10^{-3}$.}
\label{beta_series}
\end{center}
\end{figure}
\begin{figure}[htbp]
\begin{center}
\includegraphics[width=11cm]{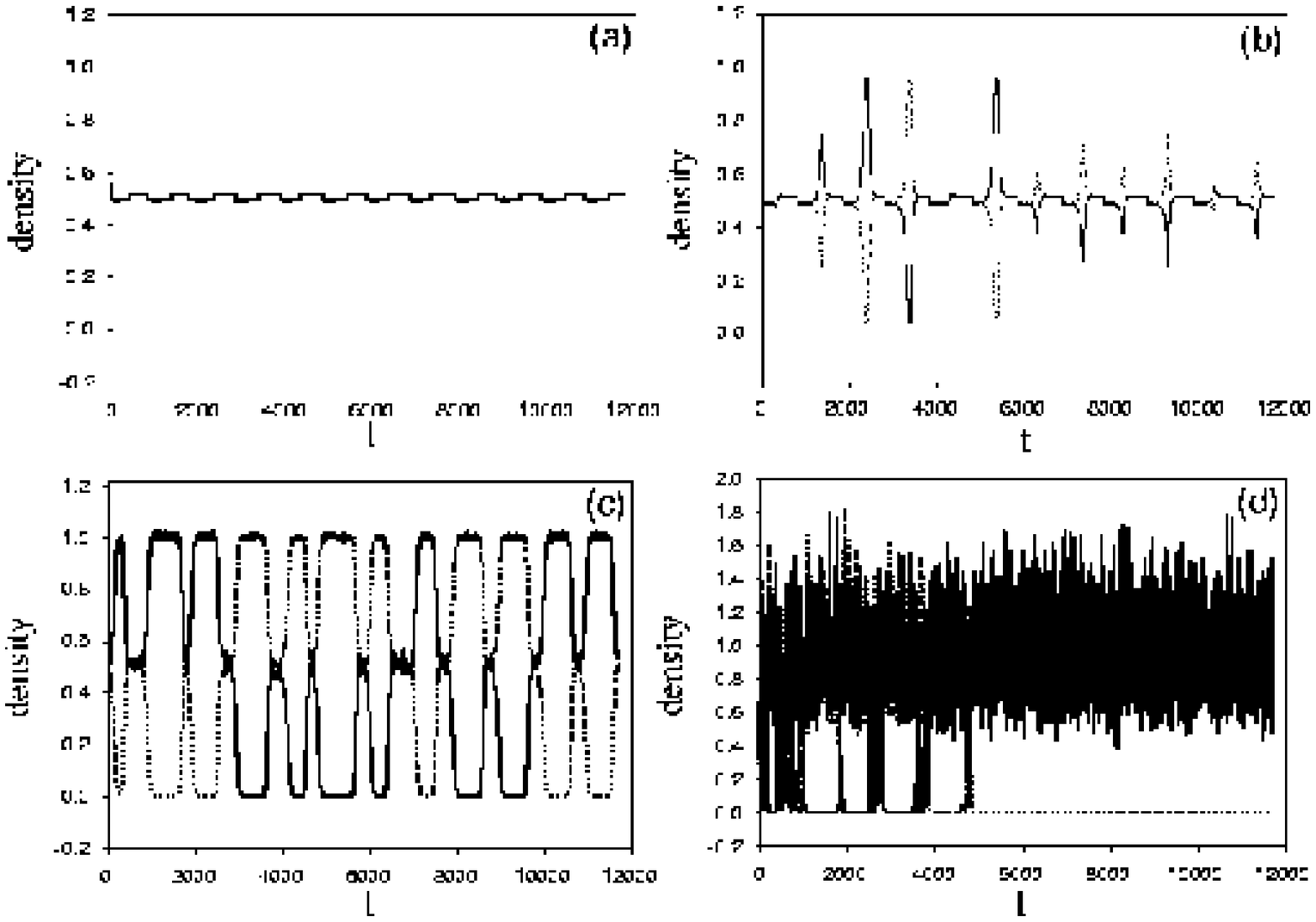}
\caption{Time evolution of both populations at different levels of
the multiplicative noise: (a) $\sigma=0$; (b) $\sigma=10^{-10}$;
(c) $\sigma=10^{-4}$; (d) $\sigma=10^{-1}$. The values of the
parameters are $\mu = 1$, $\alpha=1$, $\gamma = 10^{-1}$,
$\w0/2\pi = 10^{-3}$. The intensity of the additive noise is fixed
at the value $\sigma_\beta=1.78 \cdot 10^{-3}$. The initial values
of the two species are $x(0)=y(0)=1$.}
\label{time_series}
\end{center}
\end{figure}
Since the dynamics of the species strongly depends on the value of
the interaction parameter, we initially analyze the time evolution
of $\beta(t)$ for different levels of the additive noise
$\sigma_\beta$. To obtain the time series for the two species, we
set in Equation (\ref{LotVol}) $\alpha_1=\alpha_2=\alpha $,
$\beta_1(t)=\beta_2(t)=\beta(t)$. Depending on the value of the
multiplicative noise intensity we obtain: (i) a periodical
behavior of $\beta(t)$ in the coexistence region (see Fig. 2a);
(ii) the same behavior of Fig. 2a, slightly perturbed by the noise
(see Fig. 2b); (iii) a quasi-periodical behavior of the
interaction parameter jumping between the two values
$\beta=0.94<1$ and $\beta=1.04>1$, respectively corresponding to
left side well (coexistence regime) and right side well (exclusion
regime) of the potential shown in Fig. 1; and finally (iv) a loss
of coherence and a dynamical behavior strongly controlled by the
noise (Fig. 2d). We note in Fig. 2c synchronization of noise with
driving periodical force~\cite{Gam,Man}, the typical signature of
stochastic resonance that should appear in real ecosystems,
because of geological cause and the environmental
noise~\cite{Ben_etal_All_etal}. The dynamics of the two species is
analyzed by fixing the additive noise intensity at the value
$\sigma_\beta=1.78\cdot 10^{-3}$, corresponding to a competition
regime between the two species periodically switched from
coexistence to exclusion. The temporal series of the two species
are obtained for different values of the multiplicative noise
intensity $\sigma=\sigma_x=\sigma_y$. The initial values of the
two species are x(0) = y(0) = 1. In Fig. 3, we report the time
series of the two species densities for different values of the
multiplicative noise. For $\sigma\sim 0$ (see Figs. 3a), a regime
of coexistence with correlated oscillations between the two
species is observed. Increasing the intensity of the
multiplicative noise anti-correlated oscillations appear
characterized by a larger amplitude with periodical random
inversions of populations (see Fig. 3b-3c). For higher levels of
the multiplicative noise a degradation of the signal and a loss of
coherence of the temporal series for the species appears (see Fig.
3d). These series indicate the presence of stochastic resonance
(SR): because of a bistable potential modulated by a weak periodic
force, the response of the system may be enhanced by the presence
of the noise and a periodicity appears. We investigate the
presence of SR by considering $(x-y)^2$, the squared difference of
population densities. Fig. 4 shows the SNR of this quantity as a
function of the multiplicative noise intensity $\sigma$, for
$\sigma_\beta=1.78 \cdot 10^{-3}$. We note that dynamics of
$(x-y)$ is mainly affected by the multiplicative noise, as we can
see from Equations (1), (2). A maximum at $\sigma=10^{-4}$ is
present. The above analysis makes it clear the role of the two
noise sources: the additive noise determines the conditions for
the different dynamical regimes of the two species, and the
multiplicative noise produces a coherent response of the system by
a mechanism of symmetry breaking of the dynamical evolution of the
ecosystem.
\begin{figure}[htbp]
\begin{center}
\includegraphics[width=10cm]{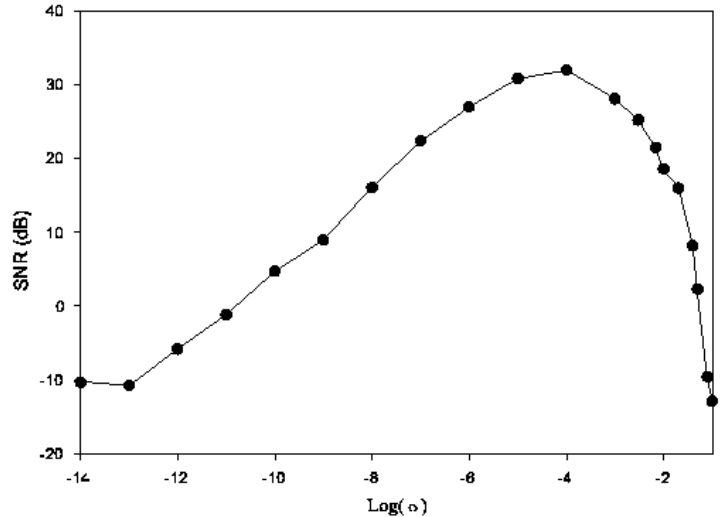}
\caption{Log-Log plot of SNR as a function of the multiplicative
noise intensity. The SNR corresponds to the squared difference of
population densities $(x - y)^2$. The values of the parameters are
the same as Figure \ref{time_series}.} \label{snr}
\end{center}
\end{figure}
\subsection{Noise-delayed extinction}\label{ss:2.2}

We now consider the mean extinction time of one species as a
function of the additive noise intensity $\sigma_\beta $, by
fixing a low value of multiplicative noise in such a way that the
system is far enough from the SR
regime~\cite{SpaFiaVal1,ValFiaSpa1}. We are not interested in the
coherent behavior of the ecological system, but we are focused on
the effect of the additive noise on the average extinction time of
the species. In Fig. 5 the usual initial condition,
$\beta(0)=0.94$, is fixed. We note that for $\sigma_\beta=0$ the
ecosystem is in the coexistence regime; that is, the deterministic
extinction time of both species is infinite. By introducing noise
causes exclusion to take place and a finite mean extinction time
(MET) appears. By varying the intensity of the additive noise in
Equation \ref{beta_eq} we obtain, of course, a variation of the
average extinction time. The delayed extinction is obtained for
noise intensities ranging from the intermediate regime (2 in Fig.
5a) to the coexistence regime obtained with higher values of
$\sigma$ (3 in Fig. 5a). This may mimic the behavior of real
ecosystems, where a finite mean extinction time may appear because
of the presence of a nonvanishing level of noise intensity. For
some environmental reason the noise intensity can change
considerably, as is observed in experimental data of populations
in a very long time interval~\cite{Car}. Therefore the dynamical
behavior shown in Fig. 5 should explain such physical situations,
where the variation of the environmental noise produces a delayed
extinction of some population. By increasing the noise intensity
we obtain noise-delayed extinction and the average extinction time
grows reaching a saturation value, that corresponds to a situation
in which the potential barrier is absent. We find nonmonotonic
behavior of the MET as a function of the noise intensity
$\sigma_{\beta}$, with a minimum value $\tau_{min}=40.47$ at
$\sigma_{\beta}=2.75 \cdot 10^{-3}$, which is of the same order of
magnitude of the barrier height $h$ (see Fig. 5a). The Kramers
time corresponding to this noise intensity is $\tau_k=41.6$, which
is approximately equal to $\tau_{min}$.
\begin{figure}[htbp]
\begin{center}
\includegraphics[width=11cm]{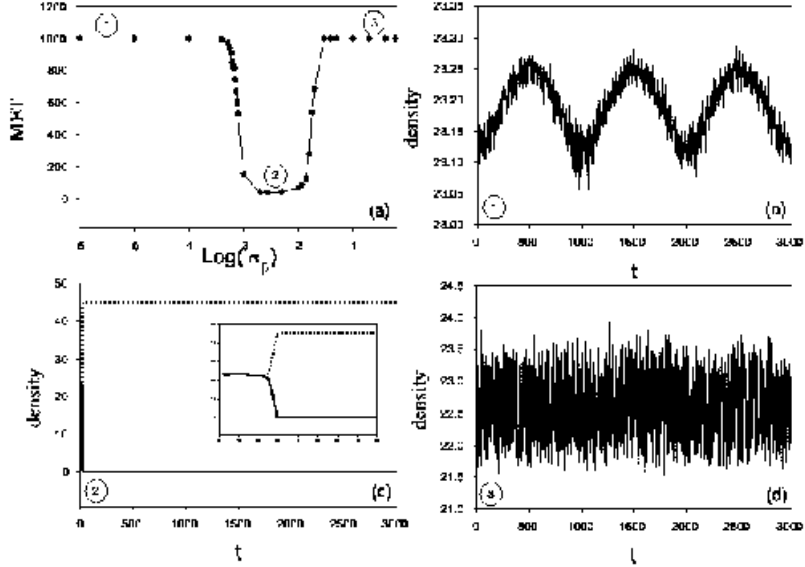}
\caption{(a) Mean extinction time of one species as a function of
the noise intensity $\sigma_\beta$. Time evolution of both species
for different levels of additive noise: (b)
$\sigma_\beta=10^{-4}$, (c) $\sigma_\beta=2 \cdot 10^{-3}$, (d)
$\sigma_\beta=10^{-1}$. The values of the parameters are $\mu =
1$, $\alpha=45$, $\gamma = 10^{-1}$, $\w0/2\pi = 10^{-3}$. The
intensity of the multiplicative noise is fixed at the value
$\sigma=10^{-9}$. The initial values of the two species are
$x(0)=y(0)=1$.} \label{met}
\end{center}
\end{figure}
This result is due to the noise-driven dynamics. In fact, for a
low value of noise intensities, the average time to overcome the
potential barrier is very high; that is, Kramers times are long.
The ecosystem remains in the coexistence regime for a long time
and the extinction time is very large. For noise intensity of the
same order of magnitude of the barrier height, the system goes
toward the exclusion regime of one of two species, and the average
extinction time is approximately equal to the Kramers time. We get
the minimum value of MET. For higher values of noise intensity,
the Kramers time becomes very small, and the representative point
of the $\beta$ parameter moves between the two minima in a very
short time. In this condition the system "sees" the average value
of the interaction parameter ($\beta=0.99$), which gives a
coexistence regime. In Figs. 5b, 5c, and 5d we show the time
evolution of the ecosystem corresponding to the points $1$, $2$
and $3$ of Fig. 5a. We have a coexistence regime in points $1$ and
$3$ and an exclusion regime in point $2$.

\subsection{Colored noise}\label{S:2.3}
In real ecosystems the external random perturbations, because of
interaction with the environment, are correlated within a finite
correlation time. When the time scale of random fluctuations is
larger than the characteristic time scale of the ecosystem the
external noise cannot be considered white noise. A strongly
correlated noise, for example, emerges as the result of a coarse
graining over a hidden set of slow variables~\cite{Gam}. In this
section we report the effect of realistic noise in the dynamics of
two competing species, and specifically on the SR phenomenon in
population dynamics in the presence of exponentially correlated
noise. The dynamics of our ecosystem is described by Equations
(1), (2), and (3). For low values of the correlation time
$\tau_c$, the response of the system coincides with that obtained
with multiplicative white noise. For higher values of $\tau_c$,
the coherent response of the system and the maximum of the
signal-to-noise ratio (SNR), which are signatures of the SR
phenomenon, are shifted towards higher values of the noise
intensity. These results agree with previous theoretical and
experimental investigations of the SR phenomenon in dynamical
systems in the presence of colored noise~\cite{Gam,Man2,Han}.
However in previous studies the colored noise was additive, while
here we have two different sources of noise and only one of them
is colored.

Now in Equations (1) and (2), $\xi_i(t)$ $(i=x,y)$ are colored
noises given by the archetypal source for colored noise; that is,
exponentially correlated processes given by Ornstein-Uhlenbeck
process~\cite{Gardiner}
\begin{equation}
\frac{d\xi_i}{dt}=-\frac{1}{\tau_c}\xi_i + \frac{1}{\tau_c}
\eta_i(t) \qquad (i=x,y)
\label{colored_noise}
\end{equation}
and $\eta_i(t)$ $(i=x,y)$ are Gaussian white noises within the Ito
scheme with zero mean and correlation function $\langle
\eta_i(t)\eta_j(t')\rangle = 2\sigma \delta(t-t')\delta_{ij}$. The
correlation function of the processes of Equation
($\ref{colored_noise}$) is
\begin{equation}
\langle \xi_i(t)\xi_j(t')\rangle = \frac{\sigma}{\tau_c}
e^{-|t-t'|/\tau_c} \delta_{ij}
 \label{correlation function}
\end{equation}
and gives $2\delta(t-t')\delta_{ij}$ in the limit $\tau_c
\rightarrow 0$. Analogous to the previous case (i.e.
multiplicative white noise), the time series for the two
populations are obtained setting $\alpha_1=\alpha_2=\alpha $,
$\beta_1(t)=\beta_2(t)=\beta(t)$, and $\xi_x(t)=\xi_y(t)=\xi(t)$,
where the interaction parameter $\beta(t)$ is described by
Equations (3), (4). The optimum coherent time behavior of
$\beta(t)$ (Fig. 2c), typical of the SR phenomenon, may be used to
obtain the time series of the two species densities in the
presence of multiplicative colored noise. Therefore we follow a
procedure analogous to that applied in the case of multiplicative
white noise: we analyze the dynamics of the two species by fixing
the additive noise intensity at the value $\sigma_\beta=1.78\cdot
10^{-3}$ (see Fig. 2c), and we vary the intensity of the
multiplicative colored noise. We obtain the time series of the two
species for different values both of the multiplicative noise
intensity $\sigma=\sigma_x=\sigma_y$ and the correlation time
$\tau_c$~\cite{ValFiaSpa1,ValFiaSpa2}. In particular we
investigate the system for (a) $\tau_c < T_o$ and (b) $\tau_c
> T_o$, with $T_o$ the period of the deterministic driving force.
In the weak correlated noise regime (a), no relevant modifications
occur in the temporal series of the two species densities in
comparison with the case of multiplicative white noise. The time
evolution of the two species shows an anticorrelated behavior with
quasiperiodical oscillations with a random inversion of the
population that predominates over the other one, as in the white
noise case. For $\tau_c \simeq T_o$ some modifications occur. In
particular for $\tau_c = 2\cdot10^{3}$ the time series of the two
species densities show anticorrelated behavior with
quasiperiodical oscillations up to $\sigma=10^{-2}$ (see Fig. 6d);
that is, a delay in the coherent output of our ecosystem.
\begin{figure}[htbp]
\begin{center}
\includegraphics[width=10cm]{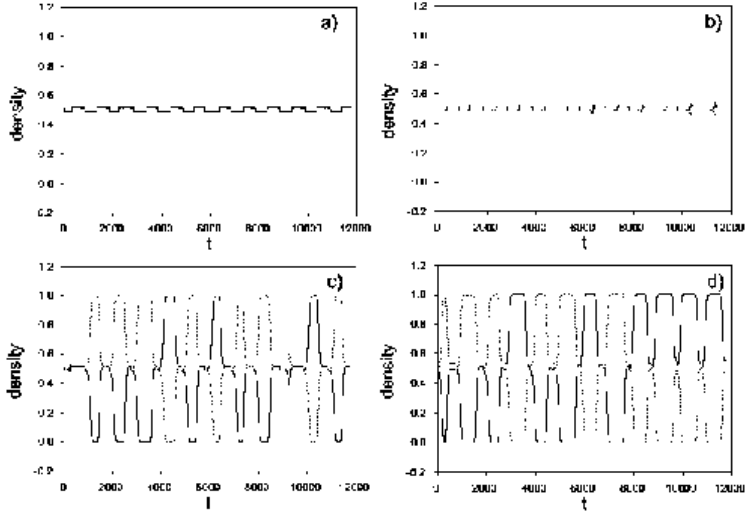}
\caption{Time evolution of both populations at different levels of
the multiplicative noise for $\tau_c=2\cdot10^3$: (a) $\sigma=0$;
(b) $\sigma=10^{-12}$; (c) $\sigma=10^{-4}$; (d) $\sigma=10^{-2}$.
The values of the parameters are $\mu = 1$, $\alpha=1$, $\gamma =
10^{-1}$, $\w0/(2\pi) = 10^{-3}$. The intensity of the additive
noise is fixed at the value $\sigma_\beta=1.78 \cdot 10^{-3}$. The
initial values are: for the two species $x(0)=y(0)=1$, for the
additive (white) noise $\beta(0)=0.94$, for the multiplicative
(colored) noise $\zeta_1(0)=\zeta_2(0)=0$.} \label{time_series_4}
\end{center}
\end{figure}
This delay will manifest itself in the behavior of the
signal-to-noise ratio (SNR) as a function of the multiplicative
noise intensity. In the strong correlated noise regime (b), a
relevant delay of the coherent time behavior of the two species is
observed. The maximum SNR is shifted toward higher values of the
multiplicative noise intensity. This shift in a Log-Log scale
grows faster than a linear function of the correlation time
$\tau_c$. The coexistence regime and the correlated oscillations
of both populations persist for a wider range of multiplicative
noise intensities. The anticorrelated behavior with
quasiperiodical oscillations appears with very high noise
intensity as the correlation time value of the multiplicative
noise is strong enough. The loss of coherence in the time
behaviors of the two species is observed at very high intensities
of the multiplicative noise. Because of the high values of the
multiplicative noise, one population extinguishes and the other
one survives at a constant density after a transient dynamics.
This dynamical behavior is typical of an ecosystem in the presence
of an absorbing barrier \cite{Ciu}. According the case of
multiplicative noise, to underline the presence of SR, we analyze
the squared difference of population densities $(x-y)^2$ for
different values of $\tau_c$.
\begin{figure}[htbp]
\begin{center}
\includegraphics[width=11cm]{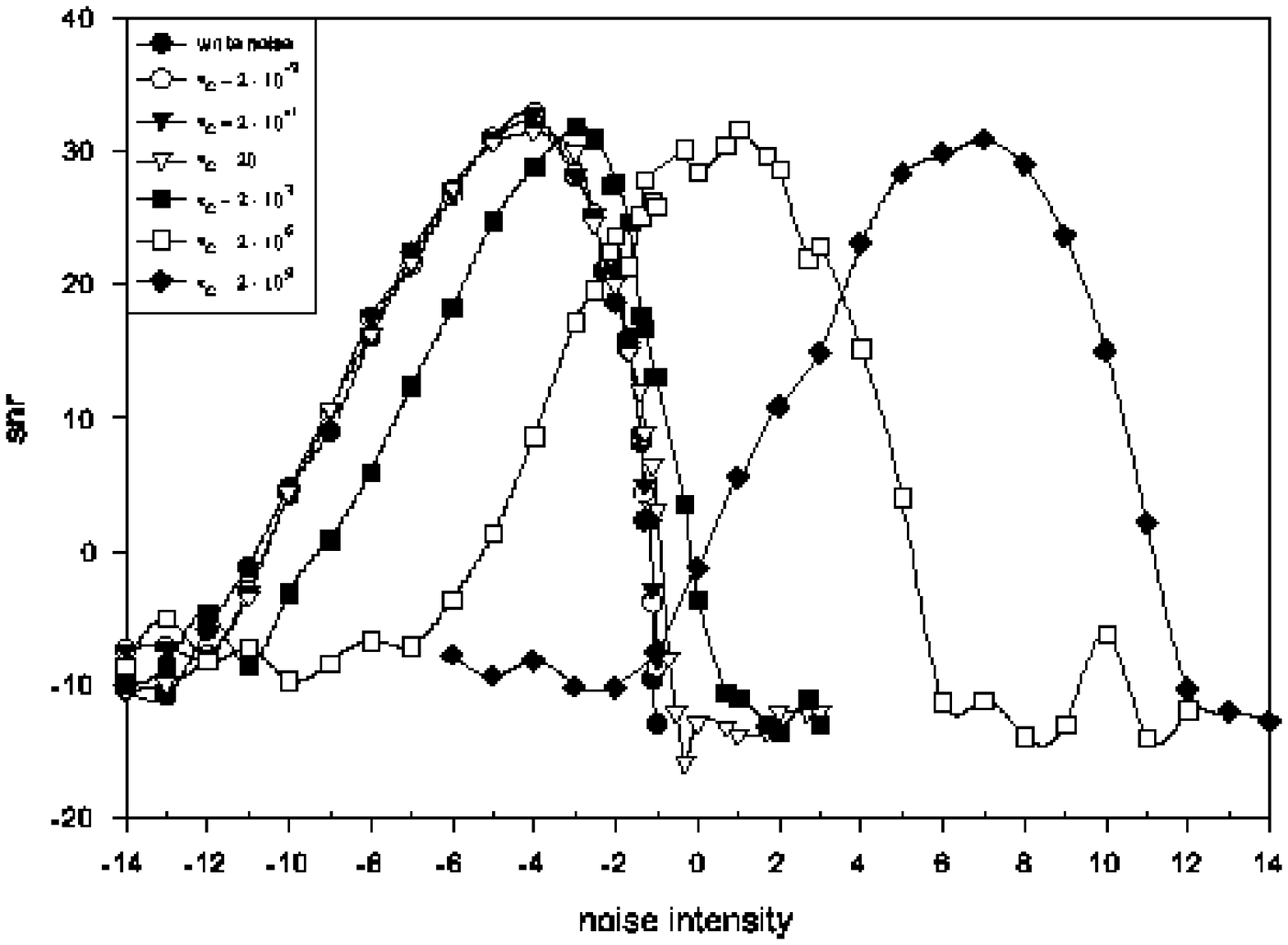}
\caption{Log-Log plot of SNR as a function of noise intensity. The
SNR is obtained for six different values of the correlation time:
$\tau_c=2\cdot10^{-2}$, $\tau_c=2\cdot10^{-1}$; $\tau_c=20$;
$\tau_c=2\cdot10^3$, $\tau_c=2\cdot10^6$; $\tau_c=2\cdot10^9$.
Moreover the signal-noise ratio for white Gaussian noise is
reported. The SNR corresponds to the squared difference of
population densities $(x - y)^2$. The values of the parameters and
the initial condirtions are the same of those used to obtain the
temporal series.} \label{snr2}
\end{center}
\end{figure}
In Fig. 7 the SNRs of this quantity are shown for
$\tau_c=0,2\cdot10^{-2},2\cdot10^{-1},20,2\cdot10^3,2\cdot10^6,2\cdot10^9$
 as a function of the multiplicative noise
intensity $\sigma$, by fixing the additive noise
intensity~\cite{ValFiaSpa2} at $\sigma_\beta=1.78 \cdot 10^{-3}$.
In each graph of this figure a maximum appears, whose position
depends on the values of $\tau_c$; that is, the most coherent
response of the system is connected with both the intensity and
the correlation time of the multiplicative noise. We see clearly
the two dynamical regimes: (a) weak correlated noise (the first
four values of $\tau_c$), (b) strong correlated noise (the last
three values of $\tau_c$). In this second regime the maximum SNR
is shifted toward higher values of multiplicative noise intensity
as in previous theoretical and experimental
studies~\cite{Gam,Man,Han}. However some differences occur.
Previous studies on the effect of colored noise on the SR
phenomenon showed that by increasing $\tau_c$ the peak of the SNR
shifts toward higher values of the noise amplitude and the maximum
decreases with a broadening of the entire curve. The shift of the
SR peak to larger noise intensities is due to the fact that
colored noise suppresses exponentially the hopping rate with
increasing noise color. In our model the colored noise is
introduced in the multiplicative noise and not in the additive one
as in usual bistable dynamical systems. The SR in the dynamics of
the interaction parameter $\beta$ induces SR phenomenon in the
dynamics of two competing
populations~\cite{ValFiaSpa1,ValFiaSpa2}. Our hopping rate in the
first SR is not affected by the \emph{``color''} of the
multiplicative noise. However, this noise is responsible for the
coherent response of the ecosystem, and therefore the presence of
color in the multiplicative noise causes the SNR peak to shift.

\subsection{Spatially extended systems}\label{S:2.4}

 To study the spatial effects due to the
presence of noise sources we consider a discrete time evolution
model, which is the discrete version of the Lotka-Volterra
equations with diffusive terms, namely a coupled map lattice
(CML)~\cite{Kaneko}

\begin{eqnarray}
x_{i,j}^{n+1}&=&\mu x_{i,j}^n (1-x_{i,j}^n-\beta^n
y_{i,j}^n)+\sqrt{\sigma_x} x_{i,j}^n X_{i,j}^n + D\sum_\gamma
(x_{\gamma}^n-x_{i,j}^n),
\label{Lotka_eq_1}\\
y_{i,j}^{n+1}&=&\mu y_{i,j}^n (1-y_{i,j}^n-\beta^n
x_{i,j}^n)+\sqrt{\sigma_y} y_{i,j}^n Y_{i,j}^n + D\sum_\gamma
(y_{\gamma}^n-y_{i,j}^n). \label{two} \label{Lotka_eq_2}
\end{eqnarray}
In Equations (\ref{Lotka_eq_1}) and (\ref{Lotka_eq_2}) $x^n_{i,j}$
and $y^n_{i,j}$ denote respectively the densities of species $x$
and species $y$ in the site $(i,j)$ at the time step $n$, $\mu$ is
proportional to the growth rate, $D$ is the diffusion constant,
$\sum_\gamma$ indicates the sum over the four nearest neighbors.
The random terms are white noise sources, modeled by independent
Gaussian variables denoted by $X^n_{i,j}$, $Y^n_{i,j}$ with zero
mean and variance unit. Here $\sigma_x$, $\sigma_y$ are the
intensities of the multiplicative noise that models the
interaction between the species and the environment. The
interaction parameter $\beta^n$ of Equations (\ref{Lotka_eq_1})
and (\ref{Lotka_eq_2}) is a stochastic process which corresponds
to the value of continuous $\beta(t)$ of Equation (\ref{beta_eq})
taken at the step $n$ and $\omega_0/2\pi=10^{-2}$.

We consider the time evolution of the spatial distribution of the
ecosystem, described by Equations  (\ref{Lotka_eq_1}) and
(\ref{Lotka_eq_2}), in the SR dynamical regime obtained for
$\sigma_\beta=2.65 \cdot 10^{-3}$. We fix the additive noise at
this value and vary the intensities of multiplicative noise.

We obtained spatio-temporal patterns of the two species for
different values of the multiplicative noise intensity $\sigma =
\sigma_x = \sigma_y$, namely $\sigma = 10^{-12}, 10^{-8}, 10^{-4},
10^{-1}$ with $\mu = 2$, $D = 0.05$, $\gamma = 1.5 \cdot 10^{-1}$,
$\omega_0/(2\pi) = 10^{-2}$, $\beta(0)=0.94$ and
$x^0_{i,j}=y^0_{i,j}=0.5$ at all sites $(i,j)$~\cite{ValFiaSpa3}.
For very low noise intensity an average correlation on the
considered lattice ($N = 100 \times 100$) between the species is
observed. For higher noise intensities an anticorrelation between
the two species is observed: the two species tend to occupy
different positions. The anticorrelation is more evident for
$\sigma = 10^{-4}$ (see Fig. 8a). Increasing the multiplicative
noise reduces the anticorrelation strongly (see Fig. 8b).

\begin{figure}[htbp]
\begin{center}
\includegraphics[width=11cm]{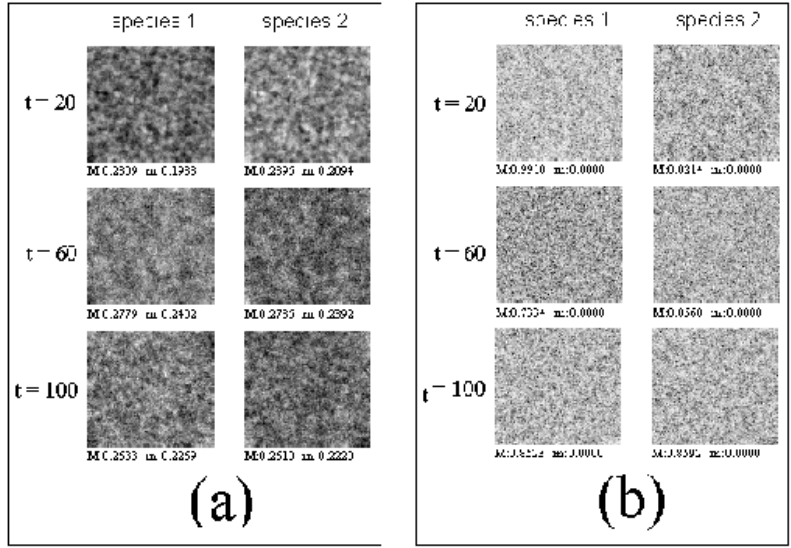}
\caption{Spatial distributions at different times for (a) $\sigma=
10^{-4}$ and (b) $\sigma= 10^{-1}$. The value of the additive
noise is fixed at $\sigma_\beta = 2.65 \cdot 10^{-3}$. The values
of the parameters are: $\mu=2$, $D=0.05$, $\gamma=1.5 \cdot
10^{-1}$, $\omega_0/(2\pi)=10^{-2}$, $N = 100\times 100$. The
initial values are $x^0_{i,j}=y^0_{i,j}=0.5$ for all sites $(i,j)$
and $\beta(0)= 0.94$.} \label{spatial2}
\end{center}
\end{figure}
To evaluate the spatial correlation between the two species for
the noise intensities considered, we calculate, at the time step
$n$, the correlation coefficient $<c^n>$ defined on the lattice
as~\cite{ValFiaSpa3}

\begin{equation}
<c^n>=\frac{cov^n_{xy}}{s^n_x s^n_y} \label{correlation}
\end{equation}
with

\begin{equation}
cov^n_{xy}=\frac{\sum_{i,j}(x^n_{i,j}-\bar{x}^n)(y^n_{i,j}-\bar{y}^n)}{N}
\label{covariance},
\end{equation}
\\
 \noindent where $\bar{x}^n$, $s^n_x$, $\bar{y}^n$,
$s^n_y$ are the mean value and the root mean square respectively
of species 1 and species 2, obtained over the whole spatial grid
at the time step $n$, $cov^n_{xy}$ is the corresponding covariance
and $N = 100\times 100$ the number of sites which compose the
grid. The behavior of the correlation coefficient $<c^n>$ as a
function of the time for different levels of the multiplicative
noise is reported in Fig. 9~\cite{ValFiaSpa3}. We observe a
nonmonotonic behavior of $<c^n>$ as a function of the
multiplicative noise intensity. For low noise intensities $\sigma
= 10^{-12}$, $<c^n>$ shows weak oscillation around $1$, that is
strong correlation between the two species. For higher levels of
the noise $\sigma = 10^{-10}$, $<c^n>$ is affected by fluctuations
and its values vary strongly as a function of the time. A further
increase of the multiplicative noise, (i.e., $\sigma = 10^{-8}$
and $\sigma = 10^{-4}$), determines an oscillation of $<c^n>$
around a negative value (i.e., anticorrelation between the two
species), with the frequency of the periodical forcing. For higher
intensities of the noise $\sigma = 10^{-1}$, the value of the
correlation coefficient $<c^n>$ increases and it vanishes for
$\sigma = 10^{+3}$. To show clearly the nonmonotonic behavior of
$<c^n>$, we calculate the time average of the correlation
coefficient $<c^n>_t$ and we report it, as a function of the
multiplicative noise intensity, in Fig. 10.
\begin{figure}[htbp]
\begin{center}
\includegraphics[width=11cm]{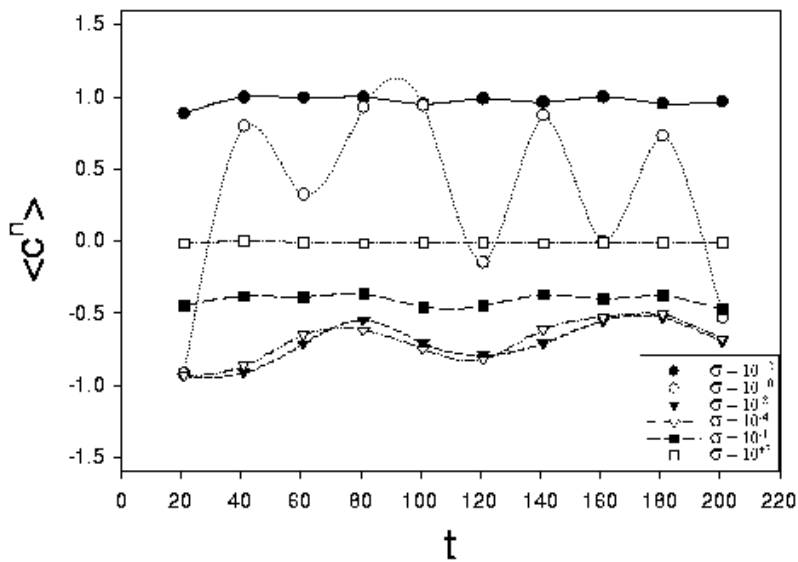}
\caption{Correlation coefficient $<c^n>$ as a function of the
time. For low levels of the multiplicative noise ($\sigma =
10^{-12}$) the species are strongly correlated and $<c^n>$ is
approximately constant. By increasing the intensity of the
multiplicative noise ($\sigma = 10^{-10}$) $<c^n>$ shows big
fluctuations. A further increase of the noise ($\sigma = 10^{-8}$,
$\sigma = 10^{-4}$) causes strong anticorrelation between the two
species with $<c^n>$ oscillating at the frequency of the
periodical forcing. For very high levels of noise, the
anticorrelation is reduced ($\sigma = 10^{-1}$) and finally it
disappears ($\sigma = 10^{+3}$); that is, the species are totally
uncorrelated.} \label{MCC}
\end{center}
\end{figure}
A clear minimum appears, which corresponds to the anticorrelated
oscillations shown in the time evolution of two competing species
in each point of our spatial grid. We note therefore the different
role of the two noise sources in the ecosystem dynamics. The
additive noise determines the conditions of the dynamical regime,
the multiplicative noise produces a coherent response of the
system~\cite{ValFiaSpa1,ValFiaSpa2}, which is responsible for the
appearance of anticorrelation behavior in the spatial patterns of
the species.

\begin{figure}[htbp]
\begin{center}
\includegraphics[width=10cm]{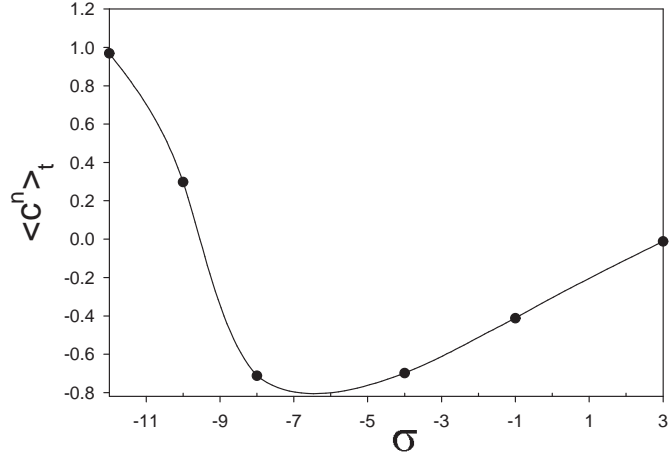}
\caption{Time average of the correlation coefficient $<c^n>_t$ as
a function of the multiplicative noise in semilog scale.}
\label{MCC_aver}
\end{center}
\end{figure}

\section{Three interacting
species}\label{S:3}

In this section we report the spatio-temporal dynamics of three
interacting species, two preys and one predator, in the presence
of multiplicative white noise and a periodical driving force. We
use the same coupled map lattice model of the previous
section~\cite{Kaneko}

\begin{eqnarray}
x_{i,j}^{n+1}&=&\mu x_{i,j}^n (1-\nu x_{i,j}^n-\beta^n
y_{i,j}^n-\gamma z_{i,j}^n)+\sqrt{\sigma_x}
x_{i,j}^n X_{i,j}^n + D\sum_\delta (x_{\delta}^n-x_{i,j}^n),\qquad\\
y_{i,j}^{n+1}&=&\mu y_{i,j}^n (1-\nu y_{i,j}^n-\beta^n
x_{i,j}^n-\gamma z_{i,j}^n)+\sqrt{\sigma_y} y_{i,j}^n Y_{i,j}^n
+D\sum_\delta (y_{\delta}^n-y_{i,j}^n),\\
z_{i,j}^{n+1}&=& \mu_z z_{i,j}^n
[-\beta_z+\gamma_z(x_{i,j}^n+y_{i,j}^n)] +
\sqrt{\sigma_z}z_{i,j}^n Z_{i,j}^n + D\sum_\delta
(z_{\delta}^n-z_{i,j}^n),\label{}
\end{eqnarray}
where $x_{i,j}^n$, $y_{i,j}^n$, and $z_{i,j}^n$ are, respectively,
the densities of preys \textit{x}, \textit{y} and of the predator
\textit{z} in the site \textit{(i,j)} at the time steps
\textit{n}. Here $\gamma$ and $\gamma_z$ are the interaction
parameters between preys and predator and D is the diffusion
coefficient. In previous equations $X$, $Y$ and $Z$ are the white
Gaussian noise variables with
 \begin{equation}
   \langle X(t)\rangle =  \langle Y(t)\rangle =  \langle Z(t)\rangle = 0,
 \end{equation}
 \begin{equation}
   \langle X(t) X(t + \tau)\rangle  = \langle Y(t) Y(t + \tau)\rangle  =
   \langle Z(t) Z(t + \tau)\rangle = \delta(\tau),
 \end{equation}
 \\
$\sigma_x=\sigma_y=\sigma_z = q$ is the noise intensity, and $\mu$
and $\mu_z$ are scale factors. $\sum_{\delta}$ indicates the sum
over the four nearest neighbors in the map lattice. The boundary
conditions have been established in such a way that no interaction
is present out of lattice. The interaction parameter $\beta$
between the two preys is a periodical function whose value, after
\textit{n} time steps, is given by

\begin{equation}
\beta(t)=1+\epsilon+\alpha cos(\w0 t), \label{beta(t)}
\end{equation}
\\
with $\epsilon=-0.01$, $\alpha=0.1$, and $\nu_0 = (\w0/2\pi) =
10^{-3}$.
\begin{figure}[htbp]
\centering{\resizebox{11cm}{!}{\includegraphics{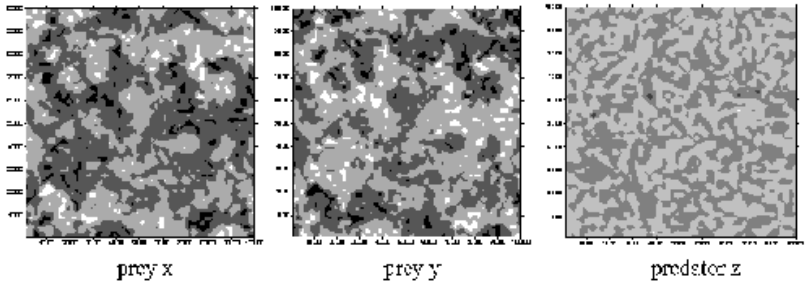}}}
\caption{Spatial patterns induced by the noise for three
interacting species (two preys and one predator) with homogeneous
initial distributions. The parameter set is: $\epsilon=-0.01$,
$\mu = 2$, $\mu_z=1$, $\nu=1$, $\beta_z = 0.01$, $\nu_0 =
(\w0/2\pi) = 10^{-3}$, $\alpha=0.1$, $\sigma_x = \sigma_y =
\sigma_z = 10^{-8}$, $D = 0.01, \gamma = 3\cdot10^{-2}$, $\gamma_z
= 2.05 \cdot 10^2$. The initial values of the uniform spatial
distribution are $x^{init}_{i,j} = y^{init}_{i,j}= 0.25$ and
$z^{init}_{i,j} = 0.10$ for all sites \textit{(i,j)}.}
\label{init_cond_1}
\end{figure}
\begin{figure}[htbp]
\centering{\resizebox{11cm}{!}{\includegraphics{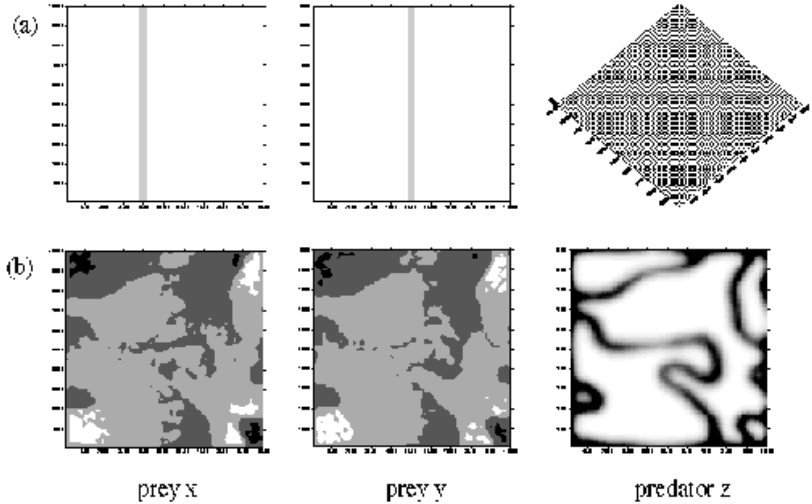}}}
\caption{Spatial patterns induced by the noise for three
interacting species (two preys and one predator) with delta-like
initial distributions of the preys and a homogeneous distribution
of the predator: (a) initial conditions, (b) spatial patterns
after 800 time steps. Here we set $\epsilon=-0.05$, $D = 0.1$,
$\sigma_x = \sigma_y = \sigma_z = 10^{-3}$ and the other
parameters are the same as in Figure ~\ref{init_cond_1}.}
\label{init_cond_2}
\end{figure}
The interaction parameter $\beta(t)$ oscillates around the
critical value $\beta_c=1$ in such a way that the dynamical regime
of Lotka-Volterra model for two competing species changes from
coexistence of the two preys ($\beta<1$) to exclusion of one of
them ($\beta>1$). We consider two different initial conditions:
(i) a homogeneous initial distribution,  and (ii) a peaked initial
distribution. In the first case we find exactly anticorrelated
spatial patterns of the two preys, while the spatial patterns of
the predator show correlations with both the spatial distributions
of the preys (see Fig. 11). The preys tend to occupy different
positions as in the case of two competing species. In the second
case, we use delta-like initial distributions for the two preys
and a homogeneous distribution for the predator. After $800$ steps
we find strongly correlated spatial patterns of the preys that
almost overlap each other. The maximum of spatial distribution of
the predator is just at the boundary of the spatial concentrations
of the preys, so that the predator surrounds the preys (see Fig.
12). The preys now tend to overlap spatially, as occurs in real
ecosystems when preys tend to defend themselves against the
predator attacks~\cite{SpaFiaVal1}.

The quantitative calculations of the site correlation coefficient
between a couple of species in the lattice have been done using
the following formula
\begin{equation}
r^n = \frac{\sum_{i,j}^N (w_{i,j}^n - \bar{w}^n) (k_{i,j}^n -
\bar{k}^n)}{\left[\sum_{i,j}^N (w_{i,j}^n - \bar{w}^n)^2
\sum_{i,j}^N (k_{i,j}^n - \bar{k}^n)^2 \right]^{1/2}}~,
 \label{r}
\end{equation}
where $N$ is the number of sites in the grid, the symbols $w^n,
k^n$ represent one of the three species $x, y, z$, and
$\bar{w}^n,\bar{k}^n$ represent the mean values of the
concentration of the species in all the lattice at the step $n$.
The two-dimensional spatial grid considered is composed by $N =
100 \times 100$ sites  in $(x,y)$ plane. The calculations have
been done for various noise intensities and at different steps of
the iteration process. To quantify our analysis, we consider only
the maximum patterns, defined as the ensemble of adjoining sites
in the lattice for which the density of the species belongs to the
interval $[3/4 \; max, max]$, where $max$ is the absolute maximum
of density in the specific grid~\cite{FiaValSpa1}. For each
spatial distribution, in a temporal step and for a given noise
intensity value, the following quantities have been evaluated
referring to the maximum pattern (MP): mean area of the various
MPs found in the lattice and spatial correlation $r$ between two
preys, and between preys and predator. The parameters used in our
simulations are as follows: $\alpha = 0.2$, $\omega_0 = \pi
10^{-3}$, $\epsilon=-0.1$ $\mu = 2$, $\nu = 1$, $\gamma = 0.03$,
$\mu_z = 0.02$, $\gamma_z = 205$, and $D = 0.1$. The noise
intensity $\sigma_x=\sigma_y=\sigma_z$ varies between $10^{-12}$
and $10^{-2}$.

\subsection{Deterministic analysis}\label{ss:3.1}

\begin{figure}[htbp]
\begin{center}
\includegraphics[height=8cm]{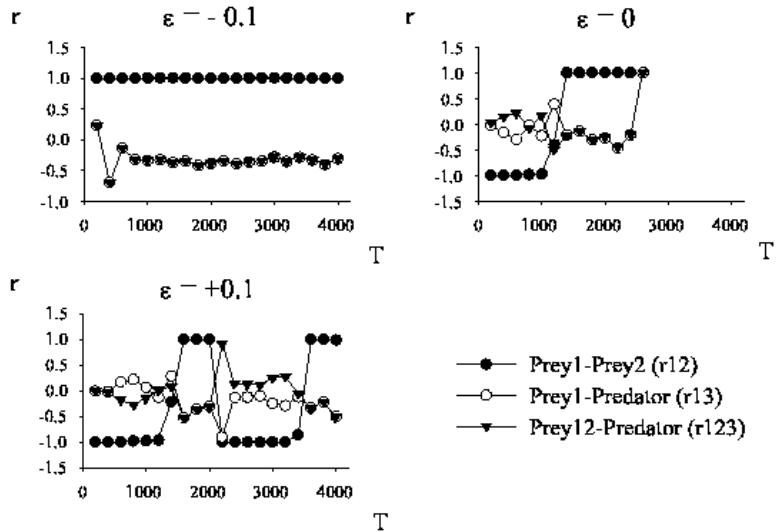}
\caption{Site correlation coefficient r in noiseless dynamics as a
function of time for different values of the parameter $\epsilon:
- 0.1, 0.,+ 0.1$. Here $\eta = 0.2$. The parameter set is: $\beta
= 1.1, q = 0, D = 0.1, \mu = 2, \nu = 1, \alpha = 0.03, \mu_z =
0.02, \gamma = 205.$ The initial conditions are random with a
Gaussian distribution, with mean values
$\bar{x}(0)=\bar{y}(0)=\bar{z}(0)= 0.25$ and variance $\sigma_o =
0.1$. Here $r_{12}, r_{13}, r_{23}$ and $r_{123}$ are respectively
the site correlations between: (i) preys, (ii) prey 1 and
predator, (iii) prey 2 and predator, and (iv) predator and both
preys.} \label{cor1}
\end{center}
\end{figure}
In the absence of noise and setting constant the value of the
interaction parameter $\beta$ we obtain: (i) for $\epsilon < 0$
($\beta < 1$) a coexistence regime of the two preys characterized
in the lattice by a strong correlation between them with the
predator lightly anticorrelated with the two preys; (ii) for
$\epsilon > 0$ ($\beta > 1$) wide exclusion zones in the lattice
(see Fig. 13), characterized by a strong anticorrelation between
preys.

By considering the periodic variation of the interaction parameter
$\beta(t)$, we obtain for $ \epsilon = 0 $, after a transient
anticorrelated behavior between preys, a coexistence regime with
strong correlation between preys that evolves toward an
homogeneous spatial distribution of all three species. For
$\epsilon > 0$, we find an oscillating behavior of the site
correlation coefficient from coexistence regime between preys
(corresponding to strong correlation) to an exclusion regime,
corresponding to strong anticorrelation. This last behavior is
prevalent. The oscillating frequency coincides with that of the
$\beta$-parameter. When $\epsilon < 0$, the two preys, after an
initial transient, remain strongly correlated for the entire time,
despite the fact that the parameter $\beta(t)$ takes values
greater than $1$ during the periodical evolution. This situation
corresponds to a coexistence regime between preys. In Fig. 13, we
report the behavior of the site correlation coefficient $r$ as a
function of time for three values of the parameter $\epsilon=-0.1,
0, 0.1$~\cite{FiaValSpa1}.

\begin{figure}[htbp]
\begin{center}
\includegraphics[height=4cm]{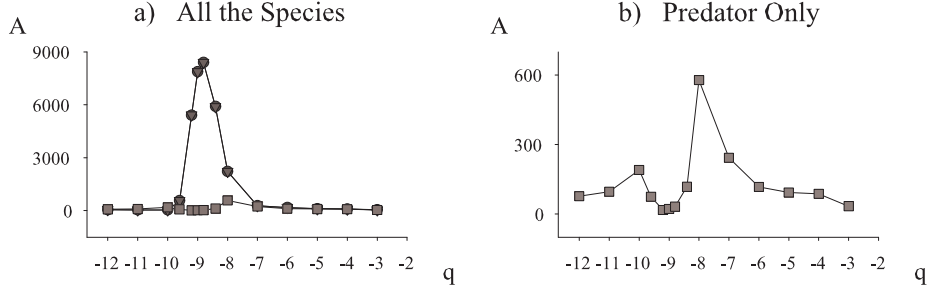}
\caption{Semi-log plot of the mean area of the maximum patterns
for all species as a function of noise intensity, at iteration
step $1400$. Here circles and triangles are related to preys,
squares to predator and $\epsilon = -0.1, \eta = 0.2$. The values
of the other parameters are the same used for Figure 13. The
initial spatial distribution is homogeneous and equal for all
species, i.e. $x_{ij}^{init}=y_{ij}^{init}=z_{ij}^{init}= 0.25$
for all sites ($i,j$).} \label{ar1}
\end{center}
\end{figure}

\subsection{Spatial patterns induced by noise}\label{ss:3.2}
To analyze the effect of the noise, we focus on the interesting
dynamical regime characterized, in absence of noise, by
coexistence between preys in all the period of $\beta$ that is,
with $\epsilon<0$.
\begin{figure}[htbp]
\begin{center}
\includegraphics[height=14cm]{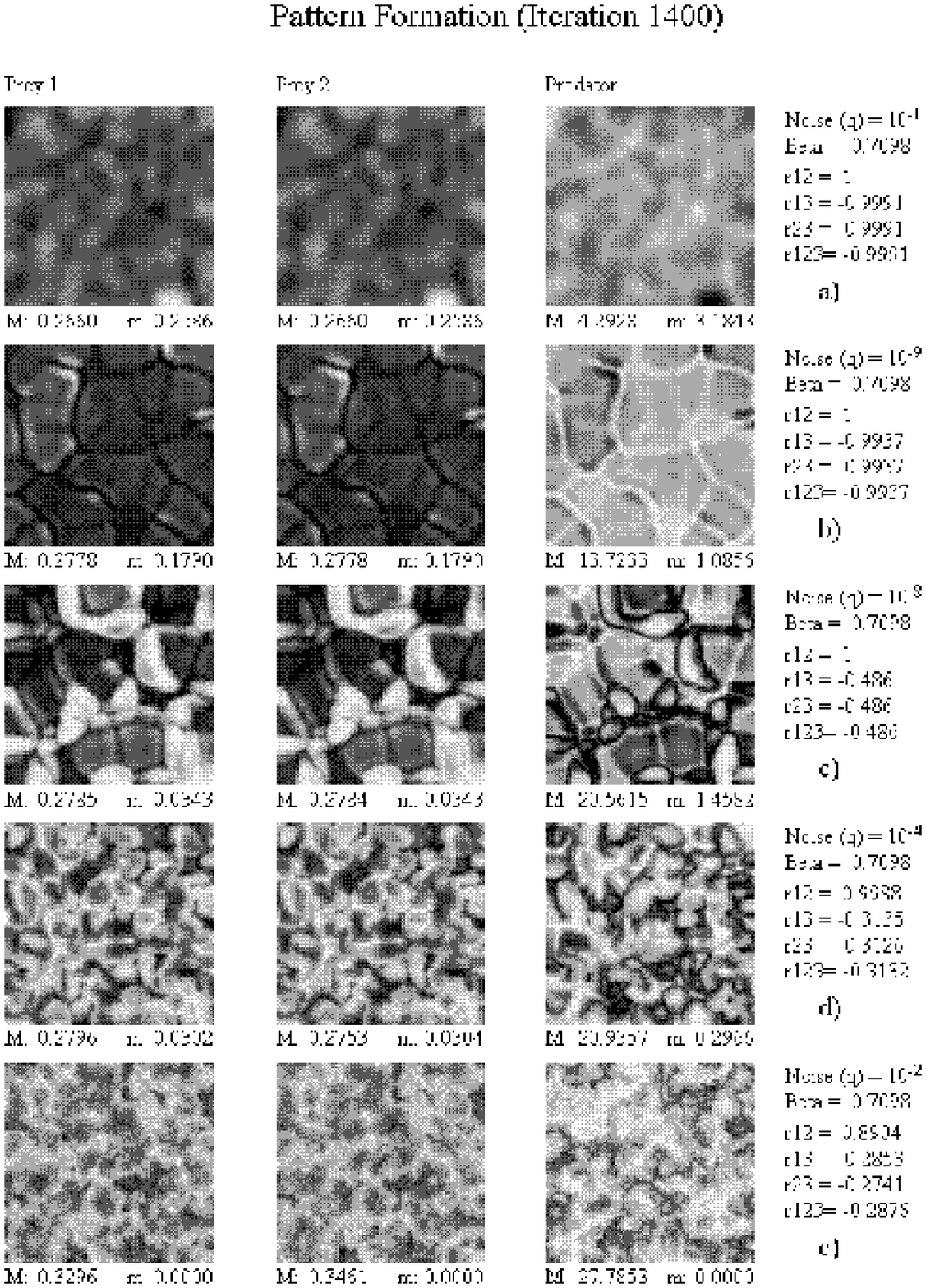}
\caption{Spatial pattern formation for preys and predator, at time
iteration $1400$ and for the following values of the noise
intensity: $q = 10^{-11}, 10^{-9},10^{-8},10^{-4},10^{-2}$. The
values of the other parameters and the homogeneous initial
distribution are the same used in Figure 14. The parameters
$r_{12}, r_{13},r_{23},r_{123}$ have the same meaning of Figure
13.} \label{pat2}
\end{center}
\end{figure}
The noise triggers the oscillating behavior of the site
correlation coefficient $r$ giving rise to periodical alternation
of coexistence and exclusion regime. Even a very small amount of
noise is able to destroy the coexistence regime periodically.
Noise is also responsible for a nonmonotonic behavior of the area
of spatial patterns, which repeats periodically in time. In Fig.
14, we report a nonmonotonic behavior of the area of the maximum
pattern as a function of noise intensity. A maximum of the area of
maximum patterns is visible for the preys at $q = 10^{-9}$ and for
the predator at $q = 10^{-8}$. The same behavior is present in the
following time steps within the first period of the interaction
parameter: 600, 800, 1200, and 1400. But at time steps $600$ and
$800$ the preys are highly uncorrelated with site correlation
coefficient $r_{12} = -1$, while at time steps $1200$ and $1400$,
the preys are highly correlated with $r_{12} = 1$. The formation
of spatial patterns appears only when the preys are highly
correlated, while large patches with clusterization of preys
appear when they are uncorrelated. This means that the coexistence
regime between preys corresponds to the appearance of spatial
patterns, while the exclusion regime corresponds to clusterization
of preys. The noise-induced pattern formation relative to the
iteration 1400 is visible in Fig. 15, where we report five
patterns of one prey and the predator for the following values of
noise intensity: $q=10^{-11}, 10^{-9},10^{-8},10^{-4},10^{-2}$.
The initial spatial distribution is homogeneous and equal for all
species; that is, $x_{ij}^{init}=y_{ij}^{init}=z_{ij}^{init}=
0.25$ for all sites ($i,j$). A spatial structure emerges with
increasing noise intensity. This spatial pattern disappears for
sufficiently large noise intensity (see Fig. 15e). As a final
investigation, we analyze the behavior of the area of the patterns
as a function of time. We observe a nonmonotonic behavior of the
area of MPs as a function of time for all values of the noise
intensity investigated. Particularly for noise intensity values
greater than $q= 10^{-7}$, this nonmonotonic behavior becomes
periodic in time with the same period of $\beta(t)$, as shown in
Fig. 16 for $q= 10^{-4}$. We note that this nonmonotonic behavior
doesn't mean that a spatial pattern appears, like that of Fig.
15b, but that a big clusterization of preys density may occur. The
maximum at $q = 10^{-4}$ of Fig. 16a in fact corresponds to large
patches of preys in the lattice investigated~\cite{FiaValSpa1}.
The various quantities, such as pattern area and correlation
coefficient, have been averaged over 200 realizations, obtaining
the mean values shown in the Figs. 14 and 16.
\begin{figure}[htbp]
\begin{center}
\includegraphics[height=4cm]{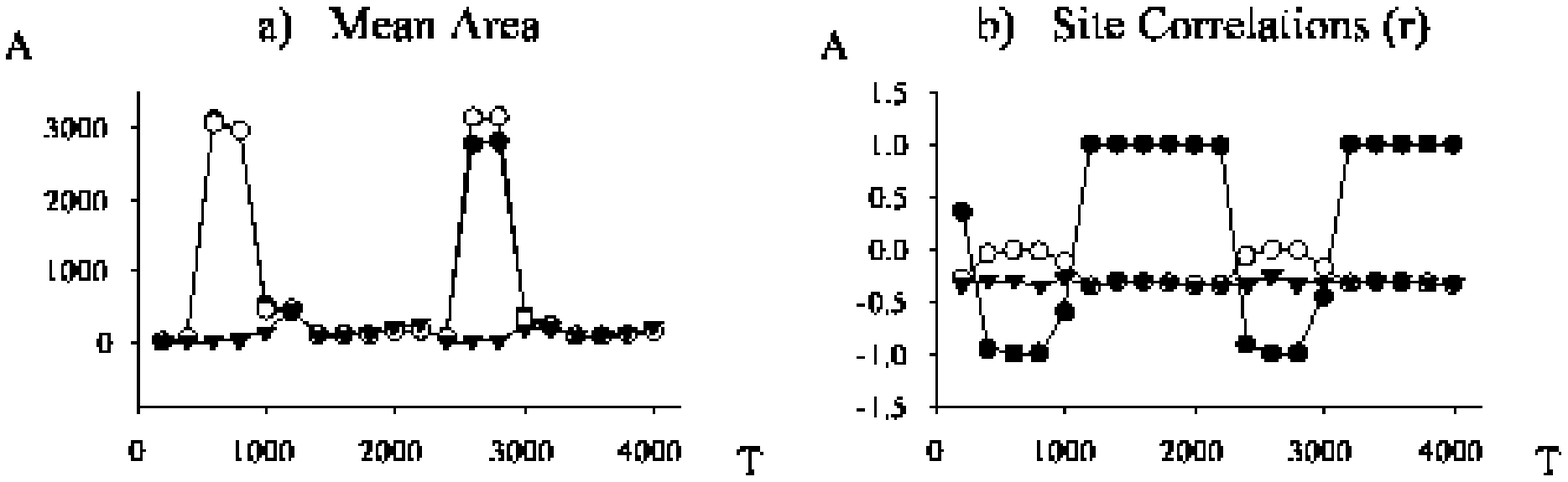}
\caption{Mean area of Maximum pattern of the three species and
relative sites correlations between preys and between preys and
predator as a function of time and $q = 10^{-4}$. (a): black and
white circles are related to preys, triangles to predator; (b)
site correlation coefficient $r_{12}$ (black circles), $r_{13}$
(white circles), and $r_{123}$ (triangles). The values of the
other parameters and the homogeneous initial distribution are the
same used in Figure 13.} \label{graf}
\end{center}
\end{figure}
The effects induced by the multiplicative noise can be summarized
as follows: (i) to break the symmetry of the coexistence regime
between the preys, producing an alternation with the exclusion
regime; (ii) to trigger the oscillating behavior of the site
correlation coefficient; and (iii) to produce a nonmonotonic
behavior of the pattern area as a function of the noise intensity
with an appearance of spatial patterns.

\section{N interacting species}\label{S:4}

In the last part of this short review we report the main results
obtained by analyzing an ecosystem composed by N interacting
species in a noisy environment in the presence of an absorbing
barrier; that is, extinction of the species
\cite{Ciu,Spa2,CirPasSpa}.

We consider an N-species generalization of the usual
Lotka-Volterra system, and the Ito stochastic differential
equation describing the dynamical evolution of the ecosystem is

\begin{equation}
d n_i(t) = \left[\left(\gamma + \frac{\epsilon}{2} \right) -
n_i(t) + \sum_{j\neq i} J_{ij}n_j(t) \right]n_i(t) dt +
\sqrt{\epsilon} n_i(t) dw_i\mbox{,} \enspace \thinspace i =
1,...,N \label{langevin}
\end{equation}
where $n_i(t) \geq 0$ is the number of elements of the
\textit{$i^{th}$} species. In Equation (\ref{langevin}), $\gamma$
is the growth parameter, the interaction matrix $J_{ij}$ modelizes
the interaction between different species ($i\neq j$), and $w_i$
is the Wiener process whose increment $dw_i$ satisfies the usual
statistical properties

\begin{equation} <dw_i(t)> \thinspace = \thinspace 0\mbox{;} \;\;\;\;
<dw_i(t)dw_j(t^{\prime})> \thinspace = \thinspace
\delta_{ij}\delta(t-t^{\prime}) dt.
\end{equation}
\\
We consider all species equivalent so that the characteristic
parameters of the ecosystem are independent of the species. The
random interaction with the environment (climate, disease,etc...)
is taken into account by introducing a multiplicative noise in the
Equation (\ref{langevin}). The solution of the dynamical Equation
(\ref{langevin}) is given by

\begin{equation} n_i(t) = \frac{n_i(0)exp\left[
\delta t +\sqrt{\epsilon} w_i(t) + \int_{0}^{t}
dt^{\prime}\sum_{j\neq i}J_{ij}n_j(t^{\prime}) \right]} {1+\gamma
n_i(0) \int_{0}^{t}dt^{\prime} exp\left[\delta t^{\prime}
+\sqrt{\epsilon} w_i(t^{\prime}) + \int_{0}^{t^{\prime}}
dt^{\prime\prime} \sum_{j\neq
i}J_{ij}n_j(t^{\prime\prime})\right]}\thinspace. \label{sol
langevin}
\end{equation}
We consider two different types of interaction between the
species: (a) a mean field approximation with a symbiotic
interaction between the species; (b) a random interaction between
the species with different types of mutual interactions:
competitive, symbiotic, and prey-predator relationship.

\subsection{Mean field approximation}\label{ss:4.1}

\noindent We consider a mean field symbiotic interaction between
the species. As a consequence, the growth parameter is
proportional to the  average species concentration,

\begin{equation}
\sum_{j\neq i}J_{ij}n_j(t) = \frac{J}{N}\sum_j n_j(t) = J
m(t)\thinspace. \label{Jm}
\end{equation}
In the limit of a large number of interacting species the
stochastic evolution of the system is given by the integral
equation

\begin{equation} M(t) =
\frac{1}{N}\sum_i\ln\left(1+ n_i(0) \int_{0}^{t} dt^{\prime} e^{J
M(t^{\prime})+\gamma t^{\prime}+\sqrt{\epsilon} w_i(t^{\prime})}
\right), \label{integral eq}
\end{equation}
where $M(t)$ is the time integral of the site population
concentration average. We introduce an approximation of this
Equation (\ref{integral eq}) which greatly simplifies the noise
affected evolution of the system and allows us to obtain
analytical results for the population dynamics. We note that in
this approximation the noise influence is taken into account in a
nonperturbative way, and that the statistical properties of the
time average process M(t) are determined asymptotically from the
statistical properties of the process $w_{max}(t)$ =
$\sup_{0<t^{\prime}<t} w(t^{\prime})$, where $w$ is the Wiener
process.
 Starting from the following approximated integral
equation for $M(t)$

\begin{equation} M(t) \simeq
\frac{1}{N}\sum_i\ln\left(1 + n_i(0) e^{\sqrt{\epsilon} w_{max_i}}
\int_{0}^{t} dt^{\prime} e^{J M(t^{\prime})+\gamma
t^{\prime}}\right) , \label{M(t) approx}
\end{equation}
it is possible to analyze the role of the noise on the
stability-instability transition in three different regimes of the
nonlinear relaxation of the system: (i) toward the equilibrium
population ($\gamma>0$); (ii) toward the absorbing barrier
($\gamma<0$); (iii) at the critical point ($\gamma=0$).
Specifically at the critical point we obtain for the time average
process $M(t)/t$ as a dominant asymptotic behavior in the
stability region (namely when $J < 1$)

\begin{equation}
\frac{M(t)}{t} \simeq \left( \frac{1}{1-J} \right) \sqrt{\frac{2
\epsilon}{\pi}}\frac{1}{\sqrt{t}},
\end{equation}
and in the instability region (namely when $J>1$)

\begin{equation}
\frac{M(t)}{t} \simeq e^{\left<\ln(n_i(0))\right>}
\sqrt{\frac{2\pi}{\epsilon}} \frac{e^{\sqrt{\frac{2
\epsilon}{\pi}} \sqrt{t}}}{\sqrt{t}}
\end{equation}
\subsection{Random interaction}\label{ss:4.2}

The interaction between the species is assumed to be random and is
described by a random interaction matrix $J_{ij}$, whose elements
are independently distributed according to a Gaussian distribution

\begin{equation}
P(J_{ij}) = \frac{1}{\sqrt{2\pi\sigma^{2}_J}}
exp\left[-\frac{J^{2}_{ij}}{2\sigma^{2}_J}\right] \mbox{,} \,\,\,
\sigma^{2}_J = \frac{J^2}{N} \thinspace , \label{P(J)}
\end{equation}
where $J$ is the interaction strength and

\begin{equation}
<J_{ij}> = 0 \mbox{,} \,\,\,\,   \mbox{} \ \ \ <J_{ij} J_{ji}> =
0.
\end{equation}
With this asymmetric interaction matrix, our ecosystem contains
$50\%$ prey-predator interactions (namely $J_{ij}<0$ and
$J_{ji}>0$), $25\%$ competitive interactions ($J_{ij}<0$ and
$J_{ji}<0$), and $25\%$ symbiotic interactions ($J_{ij}>0$ and
$J_{ji}>0$). The initial values of the populations $n_i(0)$ have
also Gaussian distribution

\begin{equation}
P(n) = \frac{1}{\sqrt{2\pi\sigma^{2}_n}} exp\left[-\frac{(n -
<n>)^2}{2\sigma^{2}_n}\right]\mbox{,} \,\,\,\, \sigma^{2}_n =
0.01\mbox{,}\,\,  \mbox{and} <n> = 1. \label{P(n)}
\end{equation}
\\
The strength of interaction between the species $J$ determines two
different dynamical behaviors of the ecosystem. Above a critical
value $J_c$, the system is unstable; this means that at least one
of the populations diverges. Below the critical interaction
strength, the system is stable and asymptotically reaches an
equilibrium state. For our ecosystem this critical value is
approximately $J = 1.1$. The equilibrium values of the populations
depend both on their initial values and on the interaction matrix.
If we consider a quenched random interaction matrix, the ecosystem
has a great number of equilibrium configurations, each one with
its attraction basin. For vanishing noise ($\epsilon = 0$), the
steady state solutions of Equation (\ref{langevin}) are obtained
by the fixed-point equation

\begin{equation}
(\gamma - n_i + h_i) n_i = 0 \label{fixpoint}
\end{equation}
where

\begin{equation}
h_i = \sum_j J_{ij} n_j(t) \label{local field}
\end{equation}
\\
is the local field. For a large number of interacting species, we
can assume that the local field $h_i$ is Gaussian with zero mean
and variance $\sigma^{2}_{h_i} = <h^{2}_i> = J^2 <n^{2}_i>$

\begin{equation}
P(h_i) = \frac{1}{\sqrt{2\pi\sigma^{2}_{h_i}}}
exp\left[-\frac{h^{2}_{i}}{2\sigma^{2}_{h_i}}\right] .
\label{P(hi)}
\end{equation}
The solutions of Equation (\ref{fixpoint}) are

\begin{equation}
n_i = 0\mbox{, i. e. extinction}
\end{equation}
and
\begin{equation}
n_i = (\gamma + h_i)\Theta(\gamma + h_i)\mbox{,} \,\,\, n_i>0,
\end{equation}
\\
where $\Theta$ is the Heaviside unit step function. From this
equation and applying the self-consistent condition, we can
calculate the steady state average population and its variance.
Specifically, we have

\begin{eqnarray}
<n_i> & = & \left<(\gamma
+ h_i) \Theta(\gamma + h_i)\right> = \nonumber \\
& = &\frac{1}{\sqrt{2\pi\sigma^{2}_{h_i}}} \left[\sigma^{2}_{h_i}
exp\left[\frac{\gamma^2}{2 \sigma^{2}_{h_i}}\right] +
\frac{\gamma\sqrt{2 \sigma^{2}_{h_i}\pi}}{2} \left(1 + erf
\left(\frac{\gamma}{\sqrt{2 \sigma^{2}_{h_i}}}
\right)\right)\right],\qquad \label{average}
\end{eqnarray}
and

\begin{eqnarray}
<n^{2}_i> & = &\left<(\gamma +
h_i)^2 \Theta^{2}(\gamma + h_i) \right> = \nonumber \\
& = & \left[\left(\frac{\gamma^2 + \sigma^{2}_{h_i}}{2} \right)
\left(1 + erf \left(\frac{\gamma}{\sqrt{2\sigma^{2}_{h_i}}}
\right)\right) + \frac{\gamma}{2}
\sqrt{\frac{2\sigma^{2}_{h_i}}{\pi}} exp\left[\frac{\gamma^2}{2
\sigma^{2}_{h_i}}\right]\right].\qquad \label{variance}
\end{eqnarray}
\\
\noindent For an interaction strength $J = 1$ and an intrinsic
growth parameter $\gamma = 1$ we obtain: $<n_i> = 1.4387,
<n^{2}_i> = 4.514,$ and $\sigma^{2}_{n_i} = 2.44$. These values
agree with that obtained from numerical simulation of Equation
~(\ref{langevin}). The choice of this particular value for the
interaction strength, based on a preliminary investigation on the
stability-instability transition of the ecosystem, ensures us that
the ecosystem is stable. The stationary probability distribution
of the populations is the sum of a delta function and a truncated
Gaussian

\begin{equation}
P(n_i) = n_{e_i} \delta (n_i) + \Theta(n_i)
\frac{exp\left[-\frac{(n_i - n_{io})^2} {2 J^2
\sigma^{2}_{n_i}}\right]} {\sqrt{2 \pi J^2 \sigma^{2}_{n_i}}}.
\label{P(ni)_zeronoise}
\end{equation}
\\
The stationary probability distribution of the population
densities has been obtained, without the extinct species, in
comparison with the computer simulations for systems with N = 1000
species and for an interaction strength $J = 1$, and for $\gamma =
1$~\cite{CirPasSpa}.

Now we focus on the statistical properties of the time integral of
the {\em $i^{th}$} population $N_i(t)$

\begin{equation} N_i(t) = \int_{0}^{t} dt^{\prime} n_i(t^{\prime}),
\label{Ni(t)}
\end{equation}
\\
in the asymptotic regime. From Equation (\ref{sol langevin}) we
have

\begin{equation} N_i(t) =
\ln\left[1+ n_i(0) \int_{0}^{t} dt^{\prime} exp\left[\gamma
t^{\prime}+\sqrt{\epsilon} w_i(t^{\prime}) + \sum_{J \neq i}
J_{ij} N_j(t^{\prime})\right]\right] ,
\label{Ni Integr. Equation}
\end{equation}
\\
In Equation (\ref{Ni Integr. Equation }) the term $\sum_j
J_{ij}N_j$ gives the influence of other species on the
differential growth rate of the time integral of the {\em
$i^{th}$} population and represents a local field acting on the
{\em $i^{th}$} population~\cite{Ciu,CirPasSpa,MePaVir}

\begin{equation} h_i = \sum_j J_{ij} N_j(t) = J \eta_i.
\label{N_localfield}
\end{equation}
\\
We  use the same approximation of the Equation (\ref{M(t) approx})
and, after differentiating, we get the asymptotic solution of
Equation (\ref{Ni Integr. Equation })

\begin{equation} N_i(t) \simeq
 \ln\left[ n_i(o) e^{\sqrt{\epsilon} w_{max_i}(t) +
J \eta_{max_i}(t)} \int^t_0 dt^{\prime} e^{\gamma
t^{\prime}}\right] \label{Ni g>0}
\end{equation}
\\
where $w_{max_i}(t) = sup_{0<t^{\prime}<t}w(t^{\prime})$ and $
\eta_{max_i}(t) = sup_{0<t^{\prime}<t}\eta(t^{\prime})$. Equation
(\ref{Ni g>0}) is valid for $\gamma \geq 0$; that is, when the
system relaxes toward an equilibrium population and at the
critical point. Evaluating Equation (\ref{Ni g>0}) for $\gamma
\geq 0$, after making the ensemble average, we obtain for the time
average of the {\em $i^{th}$} population $\bar{N_i}$

\begin{equation}
\left<\bar{N_i}\right> \simeq \frac{1}{t} \left[N_w \sqrt{\epsilon
t} + \ln t + \left<\ln\left[n_i(o) \right]\right>\right] \mbox{,}
\,\,\, \gamma = 0, \label{Ni g=0}
\end{equation}
and
\begin{equation}
\left<\bar{N_i}\right> \simeq \frac{1}{t} \left[N_w \sqrt{\epsilon
t} + (\gamma + N_{\eta} + \left<\ln\left[\frac{n_i(o)}{\gamma}
\right]\right>\right] \mbox{,} \,\,\, \gamma > 0, \label{Ni
g>0_eps}
\end{equation}
\\
where $N_w$ and $N_{\eta}$ are variables with a semi-Gaussian
distribution~\cite{Ciu} and $N_{\eta}$ must be determined
self-consistently from the Equation (\ref{N_localfield}).

These asymptotic behaviors are consistent with those obtained
using a mean field approximation. We obtain in fact the typical
long time tail behavior ($t^{-1/2}$) dependence, which
characterizes nonlinear relaxation regimes when $\gamma \geq 0$.
Further, the  numerical results confirm these analytical
asymptotic behaviors of $\bar{N_i}$~\cite{Cir1}.

When the system relaxes toward the absorbing barrier ($\gamma<0$)
we get from Equation  (\ref{Ni Integr. Equation }) in the
long-time regime

\begin{equation}
\left<\bar{N_i}\right> \simeq \frac{1}{t} \left[\ln(n_i(0)) +
\ln\left[ \int^t_0 dt^{\prime} e^{\gamma t^{\prime} +
\sqrt{t^{\prime}} w_i(t^{\prime}) + j \eta_i(t^{\prime})}
\right]\right]. \label{Ni g<0}
\end{equation}
\\
In this case the time average of the {\em $i^{th}$} population
$\left<\bar{N_i}\right>$ is a functional of the local field and
the Wiener process, and it depends on the history of these two
stochastic processes. We have also analyzed the dynamics of the
ecosystem when one species is absent. Specifically, we considered
the cavity field, which is the field acting on the {\em $i^{th}$}
population when this population is absent~\cite{MePaVir}. The
probability distributions for both local and cavity field have
been obtained by simulations for a time $t = 100$ (expressed in
arbitrary units) in absence of external noise, and for two species
(namely species $1$ and $33$)~\cite{CirPasSpa}. We found that the
probability distributions of the cavity fields differ
substantially from those of local fields for the same species
unlike the spin glasses dynamics, where the two fields coincide.
The same quantities have also been calculated in the presence of
the external noise~\cite{CirPasSpa}. The effect of the external
noise is that the two fields overlap in such a way that for some
particular species they coincide. This interesting phenomenon,
which is reminiscent of the phase-transition phenomenon, was found
for some populations. The main reasons for this peculiar behavior
are (i) all the populations are positive; (ii) the particular
structure of the attraction  basins of our ecosystem; and (iii)
the initial conditions, which differ for the value of one
population, belong to different attraction basins. Some
populations, in the absence of external noise, have a dynamical
behavior such that after a long time they significantly influence
the dynamics of other species. In the presence of noise, all the
populations seem to be equivalent from the dynamical point of
view. We also found that for strong noise intensity (namely
$\epsilon =1$) all species extinguish on a long-time scale ($t
\approx 10^6$ a. u.). Whether extinction occurs for any value of
noise intensity is still an open question, because of
time-consuming numerical calculations.

\section{Conclusions}\label{S:5}

We briefly reviewed the noise-induced phenomena in population
dynamics of three different ecosystems: (i) two competing species,
(ii) three interacting species, and (iii) N-interacting species.
In the case of two competing species, we considered two sources of
white noise: a multiplicative noise and an additive noise, that
produces a random interaction parameter between the species. The
noise induces a coherent time behavior of two species, giving rise
to temporal oscillations and enhancement of the response of the
system through the stochastic resonance phenomenon. Specifically
the additive noise controls the switching between the coexistence
and the exclusion dynamical regimes, the multiplicative noise is
responsible for coherent oscillations of the two species. The SR
in the dynamics of interaction parameter $\beta$ induces SR
phenomenon in two competing species. These time behaviors are
absent in the deterministic dynamics. The noise is also
responsible for a delayed extinction that gives rise to a
nonmonotonic behavior of the average extinction time as a function
of the additive noise intensity. We evaluated the role of colored
noise and its effects on the time behavior of the two species. We
found that the multiplicative noise is responsible for periodical
oscillations of the two species densities, whose amplitude and
coherence depend on the value of the correlation time $\tau_c$.
For $\tau_c\rightarrow 0$ our results are consistent with that
obtained for the case of white noise. Moreover the coherent time
behavior of our ecosystem and the SR phenomenon are shifted
towards higher noise intensities, in agreement with previous
theoretical and experimental investigations. We note that our
model is useful to describe physical situations in which the
amplitude of the periodical driving force, because of the
temperature variations, is weak and therefore unable to produce
considerable variations of the dynamical regime of the ecosystem.
The synergetic cooperation between the nonlinearity of the system
and the random and periodical environmental driving forces
produces, therefore, a coherent time behavior of the ecosystem
investigated. We find that these noise-induced effects should be
useful to explain the time evolution of species whose dynamics are
strongly affected by the noisy
environment~\cite{Spa1,Sci99,Gar,Car}. We also analyzed the role
of the noise in spatio-temporal behaviors by using a discrete
version of Lotka-Volterra equations with diffusive terms. We found
that the noise induces spatio-temporal behaviors that are absent
in the deterministic dynamics; that is, pattern formation with the
same periodicity of the deterministic force. Moreover appearance
of temporal oscillation is observed in the correlation coefficient
between the two species and a nonmonotonic behavior of the
time-average correlation coefficient as a function of the
multiplicative noise.

We also analyzed the role of the noise on the spatio-temporal
behaviors of an ecosystem composed by three interacting species.
We found that the formation of dynamical spatial patterns occurs
with correlations which are strongly dependent on the initial
conditions. Moreover, we obtain nonmonotonic behavior of the mean
area of the maximum patterns as a function of noise intensity. We
find the same behavior for the area of the patterns as a function
of evolution time. The noise changes the dynamical regime of the
species, breaking the symmetry of the coexistence regime. In
addition, the noise produces spatial patterns and temporal
oscillations of the site correlation coefficient defined on the
lattice. Our model for spatially extended systems composed by two
and three species could be useful to explain spatio-temporal
behaviors of populations whose dynamics is strongly affected by
the noise and by the environmental physical variables; that is,
interpreting the experimental data of population dynamics strongly
affected by the noise \cite{Car,Gar}. Finally we analyzed the
nonlinear relaxation of an ecosystem composed by N interacting
species. By using an approximation of the integral equation, which
gives the stochastic evolution of the system, we obtained
analytical results that reproduce very well almost all the
transient dynamics. We investigated the role of the noise on the
stability-instability transition and on the transient dynamics.
For random interaction, we obtained asymptotic behavior for three
different nonlinear relaxation regimes. We obtain the stationary
probability distribution of the population, which is the sum of
two contributions: (i) a delta function around $n = 0$ for the
extinct species and (ii) a truncated Gaussian for the alive
species. When we switch on the external noise, an interesting
phenomenon is observed: the local and the cavity fields, whose
probability distributions are different in the absence of noise,
coincide for some populations. This phenomenon can be ascribed to
the peculiarity of the attraction basins of our ecosystem. This
model could be useful to describe plankton dynamics.
\section{Acknowledgments}\label{S:9}
This work was supported by INFM (Istituto Nazionale per la Fisica
della Materia), MIUR (Ministero dell'Istruzione, dell'Universit\`a
e della Ricerca), and INTAS Grant 01-450 (International
Association of the European Community to promote scientific
co-operation with New Independent States).

\medskip

Received for publication January 2004.

\medskip

\end{document}